\documentclass[11pt, a4paper]{article}

\usepackage{jheppub}
\usepackage{amsmath, amsfonts, amsthm, amssymb, graphicx}
\usepackage[normalem]{ulem}
\usepackage{color}
\usepackage{cancel}
\usepackage{float}
\usepackage[utf8]{inputenc}
\usepackage{lipsum}
\usepackage{centernot}
\usepackage{ulem}
\usepackage{comment}
\usepackage{mathtools}
\usepackage{enumitem}
\usepackage{feynmp}
\usepackage{comment}

\newcommand{\kah}{K\"ahler }

\renewcommand{\Im}{\text{Im}\ }
\renewcommand{\Re}{\text{Re}\ }
\newcommand{\x}{\relax\ifmmode \mathcal{X} \else $\mathcal{X}$ \fi}

\def\be{\begin{equation}}
\def\ee{\end{equation}}
\def\ben{\begin{equation*}}
\def\een{\end{equation*}}

\def\del{\partial}

\newcommand{\vol}{\relax\ifmmode \mathcal{V} \else $\mathcal{V}$ \fi}

\renewcommand{\mp}{\relax\ifmmode M_{\text{pl}} \else $M_{\text{pl}}$ \fi}
\newcommand{\vmin}{\relax\ifmmode V^{(n)}_{\text{min}}\else $V^{(n)}_{\text{min}}$ \fi}

\title{Quintessence and the Swampland: \\
\Large{The parametrically controlled regime of moduli space}}

\author[a,b]{Michele Cicoli,} 
\author[c,d]{Francesc Cunillera,} 
\author[c,d]{Antonio Padilla,} 
\author[a,b]{Francisco G. Pedro} 

\affiliation[a]{Dipartimento di Fisica e Astronomia, Università di Bologna, via Irnerio 46, 40126 Bologna, Italy} 
\affiliation[b]{INFN, Sezione di Bologna, viale Berti Pichat 6/2, 40127 Bologna, Italy} 
\affiliation[c]{School of Physics and Astronomy, University of Nottingham, Nottingham NG7 2RD, UK} 
\affiliation[d]{Nottingham Centre of Gravity, University of Nottingham, Nottingham NG7 2RD, UK}

\emailAdd{michele.cicoli@unibo.it}
\emailAdd{francesc.cunilleragarcia@nottingham.ac.uk}
\emailAdd{antonio.padilla@nottingham.ac.uk}
\emailAdd{francisco.soares@unibo.it}

\abstract{We provide evidence that slow roll is not possible in any parametrically controlled regime of the moduli space of string theory. This is proven in full generality in the asymptotic limit of the moduli space of type II and heterotic Calabi-Yau compactifications for the dilaton and any number of K\"ahler moduli. Our results suggest that in order to build quintessence into string theory one must work in the interior of moduli space where numerical, even if not parametric, control could still be achieved.}

\begin{document}

\maketitle

\section{Introduction}

Cosmological observations \cite{hep-th/9812133,astro-ph/9805201, 1807.06209} suggest that our universe has experienced at least two phases of accelerated expansion --- one at early times, usually referred to as inflation, and one in the most recent epoch,  known as dark energy.  Establishing the microscopic origin of inflation and/or dark energy, is a significant challenge for string theory phenomenology.  Indeed, it has been conjectured that the canonical form of accelerated expansion - that of a de Sitter vacuum driven by  a positive cosmological constant \cite{deSitter:1917zz} - cannot be obtained from string theory \cite{1806.08362, 1807.05193, 1810.05506}. Here our interest lies in the late time alternative to de Sitter, some times known as quintessence \cite{Peebles:1987ek, Ratra:1987rm, astro-ph/9708069}.  This is where we have a scalar field, or fields, in slow roll, yielding a dynamical theory of dark energy that is almost de Sitter, but not quite.  Such a scenario might be appealing in string theory, partly because of the difficulties in modelling late time de Sitter, and partly because they offer the possibility of a solution to the coincidence problem \cite{2105.03426}.

However, this short note is the first of two papers identifying significant challenges  to quintessence model building in string theory compactifications. Here we will prove that quintessence is not possible in the runaway region close to the boundary of moduli space, because one cannot satisfy the slow roll constraints. Related conclusions were drawn in \cite{0711.2512, 1810.09406, Ibe:2018ffn, 2005.10168, Andriot:2020vlg}. Here we build on these works by simultaneously analysing the K\"ahler-dilaton sectors, by highlighting the r\^ole of supersymmetry and by considering corrections to moduli potentials which at tree-level feature a no-scale cancellation. Further difficulties with quintessence model building in string theory were also discussed in \cite{1909.08625}. In the following we discuss the absence of a slow roll region in the runaway of type II and heterotic supergravities. For a comprehensive review of the pros-and-cons of the different theories regarding other model building aspects, see \cite{1808.08967}.

To proceed with our proof, let us approach the boundary of moduli space in a string theory compactification on some Calabi-Yau threefold, $\x$.  In the strict boundary limit, we should think of the dilaton or the volume moduli going to infinity.  As is well known, such a limit is expected to bring down an infinite tower of light states,  spoiling the effective field theory description \cite{hep-th/0605264, 1610.00010, 1802.08264, 1803.04989}. In particular, the asymptotic limits of the volume moduli bring down either Kaluza-Klein modes, consistent with a decompactification of the internal space, or winding modes, consistent with decompactification of the dual. For the asymptotic limits of the dilaton, the tower of light states is less clear --- it has been suggested that this can introduce a tower of tensionless strings and domain walls  \cite{1904.05379}.

Since the physics at the boundary is clearly not phenomenologically viable, let us relax the limit and consider what happens when the dilaton or some \kah modulus becomes arbitrarily large but finite. In this case, the effective theory corresponds to tree-level supergravity close to the boundary, where the moduli fields are in a runaway regime. We are interested in theories that might admit no four-dimensional vacuum, or potentially a supersymmetric Minkowski or AdS vacuum somewhere in the bulk of moduli space. The supersymmetric requirement serves to align our discussion with the usual swampland lore of non-supersymmetric vacua being unstable \cite{1610.01533}. It follows from the arguments given in \cite{1810.05506} that de Sitter vacua cannot arise in any parametrically controlled regime of the moduli space, and our only hope to achieve phenomenologically viable models of dark energy is to consider runaway quintessence. For example, type IIB flux compactifications feature at tree-level supersymmetric Minkowski vacua where the K\"ahler moduli are flat directions due to the underlying no-scale structure \cite{Burgess:2020qsc}. Leading order $\alpha'$ corrections, or a non-supersymmetric stabilisation of the dilaton and the complex structure moduli, can then generate this type of de Sitter runaway for large volume.
 
{In the following we will assume that there exists some mechanism to fix the complex structure moduli at tree-level in a controlled manner}, while the remaining moduli fields correspond to the axio-dilaton $S=s+i\alpha$ and the \kah moduli $T^a=\tau^a+i \theta^a$, which are identified with either two-cycles or four-cycles, depending on which is most convenient. In the runaway regime, close to the boundary,  the  dynamics of these moduli is governed by the following \kah potential in its tree-level sum separable form
\be 
K = -p\ln(\mathcal{V}) - \ln (S+\bar{S}) + K_0\ , \label{eq:kah_pot}
\ee
and a superpotential {$W$ to be specified for each supergravity}. We have denoted the internal volume  by $\vol(\tau^a)$. For any Calabi-Yau threefold $\x$, the volume is a homogeneous function of degree $3/2$ in the \kah moduli for four-cycles, or equivalently, of degree $3$ in the \kah moduli for two-cycles. Furthermore, for type IIB supergravity, we have  $p=2$, and work with four-cycles, while for type IIA and heterotic supergravities, we have $p=1$, and work with two-cycles. As already stated, the complex structure contribution to the \kah potential, $K_0$, is considered to be fixed. For further details, see \cite{hep-th/0507153, Cicoli:2013rwa}.

The Lagrangian for the scalar moduli is given  by 
\be
\mathcal{L} = K_{S\bar{S}} dS\wedge \star d\bar{S} +K_{a\bar{b}} dT^a\wedge \star dT^{\bar{b}} - V\ , \label{eq:lagrangian_gen}
\ee
where  $K_{I\bar J}=\del_I \del_{\bar J} K$ is the \kah metric for moduli space, which given equation \eqref{eq:kah_pot} is block diagonal. The scalar potential is obtained by computing 
\begin{equation} \label{V}
V=e^K (K^{I \bar J} D_I W D_{\bar J} \bar W-3 |W|^2)\ ,
\end{equation}
where $D_I W=\del_I W+W\del_I K$ is the \kah covariant derivative, and $I$ runs over all the moduli of the theory.

At the perturbative level, the axions, $\theta^a=\Im (T^a)$ and $\alpha$, do not contribute to the scalar potential and cannot play the r\^ole of runaway quintessence. Focusing instead on the saxions/moduli, $\tau^a=\Re (T^a)$ and $s$, we find that the relevant part of the Lagrangian is now given by
\be
\mathcal{L} = \frac{1}{4s^2} (\del s)^2 +K_{a\bar{b}} \del \tau^a \del \tau^b - V(s, \tau^a)\ , \label{eag2}
\ee
A necessary condition for phenomenologically viable quintessence is the existence of a slow roll regime, defined by the condition $\epsilon :=-\dot H/H^2< 1$ where $H$ is the Hubble parameter and  dot denotes differentiation with respect to proper cosmological time. For a generic multiscalar theory described by a Lagrangian  $\mathcal{L}=\frac12 Z_{IJ} \del \phi^I \del \phi^J- V(\phi^I)$ one can show in the limit of vanishing acceleration, $\ddot{\phi}^I=0$, the dynamics of the system is described by  
\be 
H^2\approx  \frac13 V \qquad\text{and}\qquad  3H \dot \phi^I +\Gamma^{I}_{J K} \dot{\phi}^J \dot{\phi}^K \approx  - Z^{IJ}\partial_{\phi^J} V, 
\ee
where $Z^{IJ}$ is the inverse of the field space metric and  $\Gamma^{I}_{J K}$ is the corresponding metric connection. These equations admit two distinct classes of slow roll solutions, whose existence and stability has been analysed in the context of dark energy in \cite{Cicoli:2020cfj}.
In this note we will focus on the cases when $3H \dot \phi^I \gg \Gamma^{I}_{J K} \dot{\phi}^J \dot{\phi}^K\rightarrow 0$, for which
\be
\dot H \approx \frac{\left(\del_{\phi^I} V \right) \dot \phi^I}{6H} \approx -\frac{(\del_{\phi^I} V) Z^{IJ} (\del_{\phi^J} V)}{18H^2}
\,,
\ee
and so
\be \epsilon =-\frac{\dot H}{H^2} \approx \frac12  \left(\del_{\phi^I} \ln V\right) {Z^{IJ}} \left( \del_{\phi^J} \ln V\right)\,.
\ee
In this regime, the requirement of quasi-de Sitter expansion implies the need for flat (multifield) scalar potentials. The other class of accelerating solutions, that we will not analyse here, forego the flatness of the  potential in exchange for large field space curvature, see \cite{Cicoli:2020cfj,Cicoli:2020noz}.\footnote{Strictly speaking the requirement is that $\Gamma^I_{JK} \dot{\phi}^J \dot{\phi}^K \gg 3 H \dot{\phi}^I$ for one of the scalars,  which can be achieved even in flat field space if, for instance, one uses polar coordinates as in \cite{Boyle:2001du}.} While promising, this avenue is not without challenges from a string model building perspective \cite{Brinkman:2022}.  

For the supergravity compactifications under consideration  \eqref{eag2}, the first slow roll parameter is given by $\epsilon=\epsilon_s+\epsilon_\vol$, where the dilaton contribution is
\begin{equation}
\epsilon_s=  s^2 \left(\del_s \ln V \right)^2\ ,
\end{equation}
and the \kah contribution is
\be
\epsilon_\vol=\frac14 (\del_{\tau^a} \ln V)   K^{a \bar b} (\del_{\tau^b} \ln V)\ ,
\ee
Throughout this paper, we will make regular use of Euler's theorem and its corollary: if $P_n$ is a homogeneous function of degree $n$ in $\tau^a$, then $$\tau^a \del_{\tau^a} \ln P_n=n\ , \qquad \tau^a \del_{\tau^a} \del_{\tau^b}\ln P_n=-\del_{\tau^b} \ln P_n\ .$$ In particular, this allows us to infer the following conditions on the derivatives of  the \kah potential  
\be
K^{a \bar b} \del_{\bar T^b} K=-2 \tau^a\ , \qquad 
 (\del_{T^a} K)   K^{a \bar b} (\del_{\bar T^b} K) =3\ ,
 \ee
where the indices run over the \kah moduli only.
The latter result holds as long as we work with four-cycle \kah moduli in type IIB supergravity, and two-cycle \kah moduli in type IIA and heterotic.
 
Let us now derive the detailed form of the tree-level scalar potential, and the corresponding slow roll parameter, for the type II and heterotic supergravities, in turn. We will see that the slow roll condition $\epsilon<1$ can never be satisfied in the runaway regions close to the boundary of moduli space. 

\section{The type IIB runaway}
\label{sec:no_go_quint}

Let us begin with type IIB supergravity (for further details, see \cite{hep-th/0507153}). In this instance, at tree-level, the superpotential is linear in the dilaton $W=h_0 S + f_0$, with $h_0$ and $f_0$ set by the three-form fluxes, respectively $H_3$ and $F_3$, that stabilise the complex structure moduli supersymmetrically. Furthermore, the \kah moduli $T^a=\tau^a +i\theta^a$ that enter the \kah potential through the volume will be identified with four-cycles, so that the volume is a homogeneous function of degree $\frac32$ in the corresponding saxions, $\tau^a$. Using the form of the F-term scalar potential given by equation \eqref{V} and the \kah metric \eqref{eq:kah_pot} with $p=2$, we obtain a scalar potential 
\be
V= {e^{K_0}\over 2s\vol^2} |h_0 \bar S-f_0|^2\ ,
\ee
where $\bar S=s-i\alpha$. If $h_0\neq 0$, the imaginary part of the axio-dilaton can be stabilised at the point where $\del_\alpha V=0$, or equivalently, $\alpha=-\Im \left( \frac{f_0}{h_0}\right)$, so that the scalar potential becomes
\be \label{potIIB}
V(s,\vol) = {e^{K_0}\over 2s\vol^2}|h_0|^2 \left(s- \Re \left(\frac{f_0}{h_0}\right)\right)^2 \ .
\ee
It follows that
\be
\epsilon_s=\left( \frac{ s+ \Re \left(\frac{f_0}{h_0}\right) }{  s - \Re \left(\frac{f_0}{h_0}\right)} \right)^2
, \qquad \epsilon_\vol=3\ ,
\ee
defined whenever $V \neq 0 $. We immediately infer that $\epsilon=\epsilon_s+\epsilon_\vol \geq 3$, ruling out slow roll in the runaway regime. Notice that vacua with $h_0=0$ feature exactly $\epsilon_s=1\,\Rightarrow\,\epsilon=4$.

For type IIB supergravity, if $h_0\neq 0$, it is possible to further stabilise the dilaton at the location of the supersymmetric minimum, $s=\Re \left(\frac{f_0}{h_0}\right)$,  where the leading order potential for the \kah moduli vanishes. When this happens, we should consider additional perturbative and non-perturbative corrections to the potential. In order to be compatible with a robust large volume expansion, these go as \cite{Burgess:2020qsc, AbdusSalam:2020ywo, Cicoli:2021rub}
\be \label{pot}
V = {\mathcal{A}\over \vol^{2+p}} + {\mathcal{B}\, e^{-f}\over \vol^{2+q}}+\frac{\mathcal{C}}{\vol^{2+r} g^n}\ldots\ ,
\ee
where $\mathcal{A}$, $\mathcal{B}$ and $\mathcal{C}$ will in general depend on the stabilised dilaton and complex structure moduli, while  $f>0$ and $g>0$ are homogeneous functions  of degree one in the $\tau^a$. In particular, $f$  corresponds to the dominant saxion in the non-perturbative expansion. In order for us to trust the large volume expansion, we require each of these terms to scale away more quickly than the leading order $1/\vol^2$ term, which happened to vanish once the dilaton was stabilised {at its supersymmetric minimum}. This implies that   $p> 0$, $ q \geq 0$, and $r+\frac{2n}{3}>0$, suggesting that the volume direction is made even steeper and the volume modulus will not slow roll in this case either. To see this explicitly, we compute the slow roll parameter in the asymptotic regime where only one of the terms dominates, in accordance with our definition of the boundary of moduli space. If $V \sim  {\mathcal{A}\over \vol^{2+p}}$, we find that $\epsilon \sim 3 \left(1+\frac{p}{2}\right)^2>3$.  In contrast, if $V \sim  {\mathcal{B}\, e^{-f}\over \vol^{2+q}}$, we   find  that 
\be
\epsilon=3 \left(1+\frac{q}{2}\right)^2 +f   (2+q)+\frac14 K^{a \bar b} f_a f_b  \geq  3\ ,
\ee
where the inequality follows from the fact that $f, q \geq 0$ and $K^{a \bar b}$ being a symmetric matrix with all positive eigenvalues.  Finally, if $V \sim \frac{\mathcal{C}}{\vol^{2+r} g^n}$ we find that 
\be
\epsilon=3 \left(1+\frac{r}{2}\right)^2 +n  (2+r)+\frac{n^2}{4} K^{a \bar b} (\ln g)_a (\ln g)_b  \geq  3\ .
\ee
In each  case we see that slow roll is impossible in this asymptotic region, where a single one of the corrections in equation \eqref{pot} dominates. It is interesting to note that leading order supersymmetry, imposed for the sake of the stability of the compactification, only aggravates the problem, by cancelling off the $1/\vol^2$ term that would otherwise dominate the potential at the boundary of moduli space and replacing it with a steeper term.

Alternatively, we could consider a non-supersymmetric stabilisation of the complex structure sector. This would induce a correction to \eqref{potIIB} of the form $\lambda/s\vol^2$, where $\lambda $ is a positive constant proportional to the F-terms of the complex structure moduli. The vacuum expectation value of the dilaton is then shifted to a non-supersymmetric value leaving a $1/\vol^2$ runaway for the volume mode, which is too steep to give rise to slow roll. Note further that if we move into the bulk of moduli space, we might hope to stabilise the volume at some fixed value and achieve slow roll along some other saxion direction, through the last term in \eqref{pot}. This is indeed possible, although such a scenario runs into further difficulties associated with a light gravitino \cite{Cicoli:2022}. It is also possible that interference between two terms in equation \eqref{pot} can give rise to a (short) field range for the volume where $\epsilon< 1$. This avenue is likely to involve considerable tuning and, by definition, lies in the bulk of moduli space and therefore will not be analysed in the present paper.

\section{The type IIA runaway}

We now turn our attention to the type IIA runaway  (for further details, see \cite{hep-th/0507153}). The superpotential now contains RR fluxes $\left(e_0, e_a, q^a\right)$, the (3,0)-component of the $H_3$ flux, $h_0$, and the Romans mass, $m$ \cite{Romans:1985tz}:
\be
W= e_0 + e_a T^a +{1\over 2}\mathcal{K}_{abc}q^a T^b T^c + {m\over 6}\mathcal{K}_{abc} T^a T^b T^c -i h_0 S\ .
\ee
Here the \kah moduli $T^a=v^a+i\theta^a $ are identified with two-cycles and the volume is a homogeneous function of degree $3$ in the corresponding saxions, $v^a$. In particular,  $\vol=\frac16\mathcal{K}_{abc} v^a v^b v^c$, where $\mathcal{K}_{abc}$ are the triple intersection numbers for the Calabi-Yau. Without additional fluxes or corrections to the tree-level action, the complex structure sector remains flat. To continue the discussion on the slow roll regime, we must therefore assume that some other mechanism under computational control exists to stabilise the complex structure sector, so that, at tree-level, it only enters through some constant in the \kah potential, $K_0=\int_\x \Omega\wedge\bar{\Omega}$.

The tree-level scalar potential for the type IIA theory is then given by
\be
V = {e^{K_0}\over 2s \vol}\left[4|h_0|^2s^2 + K^{a\bar b}\rho_a \bar \rho_b +4s  \Im (W \bar h_0 ) - 4\tau^a\Re (W \bar \rho_a ) +|W|^2 \right]\ ,
\ee
where 
\be
\rho_a=\del_{T^a}W=e_a +\mathcal{K}_{abc}q^b T^c+\frac{m}{2}\mathcal{K}_{abc} T^b T^c\ .
\ee
We can always absorb the vacuum expectation value of the axions into a redefinition of the fluxes. Therefore, without loss of generality, we set the axions to vanish.  With the axions gone, the superpotential can be written as
\be
W=e_0+e_a v^a+\vol (q^a \omega_a +m) -ih_0s\ ,
\ee
and
\be
\rho_a=e_a+\vol \left[ q^b ( \omega_{ab}+\omega_a \omega_b )+m\omega_a\right]\ ,
\ee
where $\omega_a= \del_{v^a} \ln \vol$ and $\omega_{ab}= \del_{v^a}\del_{v^b}  \ln \vol$. Keeping only the leading order  \kah terms at large volume, we obtain a scalar potential that goes as 
\be
V ={e^{K_0}\over 2s \vol} \left[ |h_0|^2s^2  +14 \Im (m \bar h_0) \vol s +|m|^2 \vol^2 \right] +\ldots
\ee
We now compute the slow roll parameter. Since the scalar potential is a function of $\vol/s$ to leading order in this asymptotic regime, we have that $\epsilon_\vol=3 \epsilon_s$. Furthermore, 
\be
\epsilon_s=	\left[ \frac{|h_0|^2-|m|^2 \frac{\vol^2}{s^2}}{ |h_0|^2  +14 \Im (m \bar h_0) \frac{\vol}{ s} +|m|^2 \frac{\vol^2}{s^2}  } \right]^2\ .
\ee
We remark that, in order to keep $\alpha'$ corrections from becoming important, we require\footnote{In units of $\alpha'=1$.}
\be
\vol_{\text{S}}\gg 1\ ,\label{eq:alpha_con}
\ee 
where $\vol_{\text{S}}$ is the volume of \x in the string frame. After expressing the Einstein frame volume in string frame, $\vol=s^{3/2}\vol_{\text{S}}$, we see that $\vol/s = \sqrt{s} \vol_{\text{S}}\gg 1$ to be consistent with \eqref{eq:alpha_con} and weak coupling. Thus, in the runaway regime, the ratio $\vol/s$ has to be very large and it follows that $\epsilon_s  \approx 1$, and so $\epsilon=\epsilon_s+\epsilon_\vol\approx 4$, ruling out slow roll. 

\section{The heterotic runaway}

We finish with the heterotic runaway, where the tree-level superpotential does not depend on the dilaton or the \kah moduli  (for further details see \cite{Cicoli:2013rwa}), so that $W=W_{\rm cs}$. The complex structure moduli are assumed to lie in their supersymmetric vacuum by the vanishing of the corresponding F-terms. As a result, the scalar potential becomes  a runaway in the dilaton and the volume modulus
\be
V=e^{K_0} {\left| W_{\text{cs}}\right|^2\over 2s\vol}\ .
\ee
Computing the slow roll parameters, we find that  $\epsilon_s=1$ and $\epsilon_\vol=3$, and so $\epsilon=4$.  Clearly slow roll cannot be achieved in the heterotic runaway.

\section{Conclusions}

We have found that the tree-level type II and heterotic supergravities cannot contain a slow roll region in a parametrically controlled regime. The result is proven for any number of \kah moduli. It is clear that obtaining a slow roll region requires breaking the form of the \kah potential in \eqref{eq:kah_pot}. This can be done by introducing corrections to the tree-level action. 
In the sequel to this paper \cite{Cicoli:2022}, we will step away from the runaway, by considering perturbative and non-perturbative corrections to this leading order behaviour in detail. There we find that phenomenologically viable quintessence requires  non-supersymmetric vacua. This suggests that proponents of the swampland should now object to quintessence  as vigorously as they object to de Sitter, placing them on a collision course with observations. 

\acknowledgments

We would like to thank Ivonne Zavala, Susha Parameswaran and Arthur Hebecker for useful discussions. AP was supported by an STFC consolidated grant number ST/T000732/1 and FC by a University of Nottingham studentship.

\end{document}